\documentclass[conference]{IEEEtran}
\IEEEoverridecommandlockouts

\usepackage{xcolor,soul,framed} 

\colorlet{shadecolor}{yellow}
\usepackage[pdftex]{graphicx}
\graphicspath{{../pdf/}{../jpeg/}}
\DeclareGraphicsExtensions{.pdf,.jpeg,.png}

\usepackage[cmex10]{amsmath}
\usepackage{array}
\usepackage{mdwmath}
\usepackage{mdwtab}
\usepackage{eqparbox}
\usepackage{url}

\usepackage{amssymb}
\graphicspath{ {./images/} }
\usepackage{orcidlink}
\usepackage{breqn}
\usepackage{physics}

\usepackage{tikz} 

\usepackage{float}

\begin{document}

\title{

Learning-Based Detection of Malicious Volt-VAr Control Parameters in Smart Inverters

}

\author{\IEEEauthorblockN{Ahmad Mohammad Saber\IEEEauthorrefmark{1}\orcidlink{0000-0003-3115-2384} ,
Amr~Youssef\IEEEauthorrefmark{7}\orcidlink{0000-0002-4284-8646},
Davor~Svetinovic\IEEEauthorrefmark{1}\orcidlink{0000-0002-3020-9556}, 
Hatem Zeineldin\IEEEauthorrefmark{1}\orcidlink{0000-0003-1500-1260}, and
Ehab El-Saadany\IEEEauthorrefmark{1}\orcidlink{0000-0003-0172-0686}}
\IEEEauthorblockA{\IEEEauthorrefmark{1}
 Electrical Engineering and Computer Science Department, Khalifa University, Abu Dhabi, UAE
}
\IEEEauthorblockA{\IEEEauthorrefmark{7}Concordia Institute for Information Systems Engineering,
Montreal, Quebec, Canada
}

}

\maketitle

\begin{abstract}

Distributed Volt-Var Control (VVC) is a widely used control mode of smart inverters. However, necessary VVC curve parameters are remotely communicated to the smart inverter, which opens doors for cyberattacks.  If VVC curves of an inverter are maliciously manipulated, the attacked inverter's reactive power injection will oscillate, causing undesirable voltage oscillations to manifest in the distribution system, which, in turn, threatens the system's stability. In contrast with previous works which proposed methods to mitigate the oscillations after they are already present in the system, this paper presents an intrusion detection method to detect malicious VVC curves once they are communicated to the inverter.  The proposed method utilizes a Multi-Layer Perceptron (MLP) that is trained on features extracted from only the local measurements of the inverter. After a smart inverter is equipped with the proposed method, any communicated VVC curve will be verified by the MLP once received. If the curve is found to be malicious, it will be rejected, thus preventing unwanted oscillations beforehand. Otherwise, legitimate curves will be permitted. The performance of the proposed scheme is verified using the 9-bus Canadian urban benchmark distribution system simulated in PSCAD/EMTDC environment.  Our results show that the proposed solution can accurately detect malicious VVC curves. 

\end{abstract}

\IEEEpeerreviewmaketitle

\section{Introduction}

\IEEEPARstart{S}{mart} grids are increasingly being equipped with renewable energy sources and information and communication technologies, which is revolutionizing the way modern grids are managed. 
Among the transformative technologies \textcolor{black}{required by both North American and European interconnection standards \cite{NAstandard,EUstandard}, distributed Volt-Var Control (D-VVC) has gained popularity over the last decade. D-VVC is a localized and autonomous method to control smart inverters that accompany Distributed Generators (DGs), such as \textcolor{black}{grid-connected} Photo-Voltaic (PV) systems, allowing the DGs to take part in fulfilling the grid reactive power demand \cite{VVC_paper,standard1}. }
A smart inverter operating in the VVC mode needs to receive the VVC droop control function from the operator of the grid-connected microgrid/distribution system \cite{standard2}.

Previous studies, e.g., \cite{VVC_paper,Farivar}, have shown that the proper design of the VVC curves, mainly the slopes of these curves, is vital for the system's stability. Very steep curves can result in sustained voltage magnitude oscillations in the system.
However, current inverters do not have a mechanism to distinguish between proper VVC curves and improper curves that could have been misconfigured or manipulated by a malicious entity \cite{Anna_adaptive}.
Motivated by previous successful attacks on inverter control systems \cite{Risks}, it is possible that malicious entities intentionally exploit the vulnerabilities in deployed inverters' firmware (\cite{Westerhof}), which allows them to use the remote update capability of smart inverters to communicate malevolent VVC curves \cite{Sahoo}.
It was experimentally proven that vicious manipulation of volt-var curves will result in significant voltage violations \cite{IET_paper}.
Manipulation of the VVC parameters can take the form of an insider attack, which renders the use of traditional cryptographic techniques (e.g., using a MAC or a signature scheme \cite{smart2016cryptography,9906538}) on the inverter side insufficient.
In a nationwide attack, adversaries can target groups of smart inverters, similar to how the Hawaiian utilities \textcolor{black}{were able to update} the control functions of a staggering 800,000 inverters in less than 24 hours \cite{Fairley}.
Therefore, detecting malicious curves is a crucial need.


In this paper, we present a learning-based Intrusion Detection system for VVC modes in smart inverters. 
The proposed method, which relies on only the local measurements of the inverter, employs a Multi-Layer Perceptron (MLP), which is trained offline to differentiate between legitimate curves and malicious curves that would result in stability issues in the system. 
The proposed system is completely decentralized, and all the features utilized by the MLP are extracted from the available inverter measurements. 
Once trained, the proposed scheme can be implemented within smart inverters to verify a VVC curve once received, and detect and reject malicious VVC curves. 
The proposed scheme can detect stealthy attacks that can bypass detection schemes proposed by some of the related works.

In the remainder of this paper, Section \ref{Section:related} depicts the key differences between the proposed intrusion detection system and related works.
Section \ref{Section:VVC} demonstrates how cyberattacks can target VVC function of smart inverters.
The proposed method is developed in Section \ref{Section:IDM}.
Furthermore, a case study is presented in Section \ref{Section:Case} where the proposed system is implemented to evaluate its performance. 
Section \ref{Section:flowchart} shows how the proposed method can be implemented to secure Volt-Var Control Against Setpoint Manipulation Cyberattacks.
Finally, the paper is concluded in Section \ref{Section:conclusion}.

\section{Preliminaries and Related Works}\label{Section:related}

\begin{figure}[t!]
\centering
\includegraphics[width=0.7\linewidth,keepaspectratio]{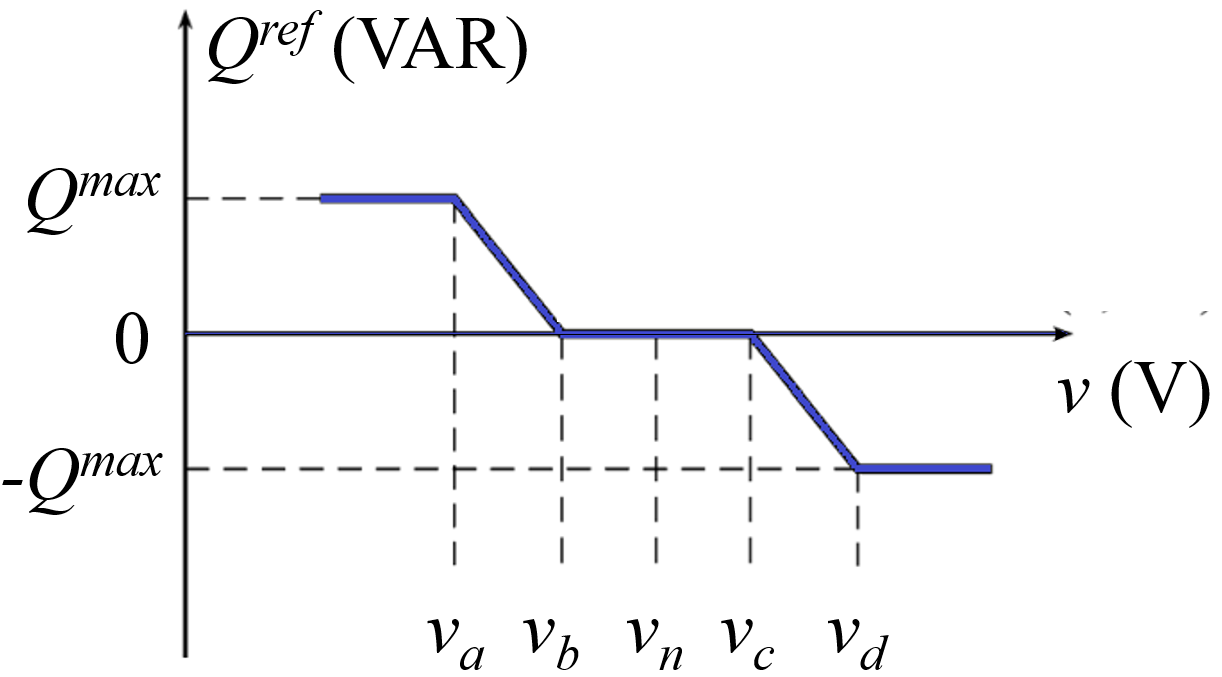}
\caption{  Illustration of distributed VVC droop curve.  The reactive power a smart inverter injects is determined based on the magnitude of the local voltage. Positive $Q$ values denote injected VAR. $v_n $ is the nominal voltage.}
\label{fig:VVC}
\end{figure}

\textcolor{black}{In centralized VVC, a centralized controller $-$typically the distribution transformer's online tap-changer$-$ coordinates reactive power ($Q$) support between itself and the downstream capacitors and DGs \cite{C_VVC_oldpaper}. This approach relies on 2-way communication links and is, therefore, highly vulnerable to cyberattacks and more prone to other communication-related problems \cite{VVC_optimizaton_attack}.
On the other side,  D-VVC offers a decentralized and autonomous means of Reactive Power Control (RPC) in smart distribution systems, where individual DGs participate in RPC based on a predetermined control function that uses only the local measurements of the DG's grid connection point.}
\textcolor{black}{
Fig. \ref{fig:VVC} shows the typical VVC droop function currently available in smart inverters \cite{VVC_paper}. In a decentralized fashion, this function, which relates the required reactive power to be injected by the inverter ($Q^{ref}$) to the locally measured voltage ($v$), relies on a group of setpoints $ \hat{v} = [v_a, v_b, v_c, v_d]$ that can be remotely adjusted by the system operator.
}

A new false-data-injection attack (FDIA) against centralized VVC was proposed in \cite{VVC_optimizaton_attack}, which can lead to significant under-voltage events.
Cybersecurity of voltage control in distribution systems with PVs was investigated \cite{Cyberattacks_v_control}.
A cyberattack detection method for PV farms was proposed in \cite{PVfarms} based on Harmonic State Space Modeling. 
A state-estimation-based centralized VVC strategy that accounts for possible cyberattacks that can manipulate DG measurements in centralized VVC schemes was proposed in \cite{SE_C_VVC}.

Impacts of cyberattacks on smart inverter settings were investigated in \cite{Olowu2020}.
The authors of \cite{IET_paper} demonstrated that if VVC curves of a smart inverter are malevolently modified, especially inverted, the inverter exhibits an abnormal behavior of generating reactive power at high voltage levels and absorbing reactive power at low voltage levels. 
More recently, in \cite{Anna_adaptive}, the authors focused on mitigating the \textcolor{black}{voltage oscillations that occur in the power system \textit{after} cyberattacks already succeed in manipulating D-VVC curves of DGs in this system.} 
In \cite{edge}, an attempt was made to develop mechanisms to detect malicious VVC curves, \textcolor{black}{which is essentially a set of restrictions on the VVC curve based on physical invariants.}  
\textcolor{black}{
The main idea was to judge newly received curve parameters based on the slope of the line connecting $v_a$ and $v_d$. 
However, this detection mechanism can be easily bypassed if individual curve parameters are manipulated, e.g., shifted, while keeping the overall slope of the line $v_a$-$v_d$ unchanged (as will be shown in Section \ref{Section:VVC}).
} 
Therefore, it can be noticed that there is still a gap in developing an intrusion detection method to accurately detect and prevent improper VVC curves, which we aim to fill in this paper. 

\section{Vulnerability of Distributed Volt-Var Control to Cyberattacks} \label{Section:VVC}

\begin{figure}[t!]
\centering
\includegraphics[width=0.95\linewidth,keepaspectratio]{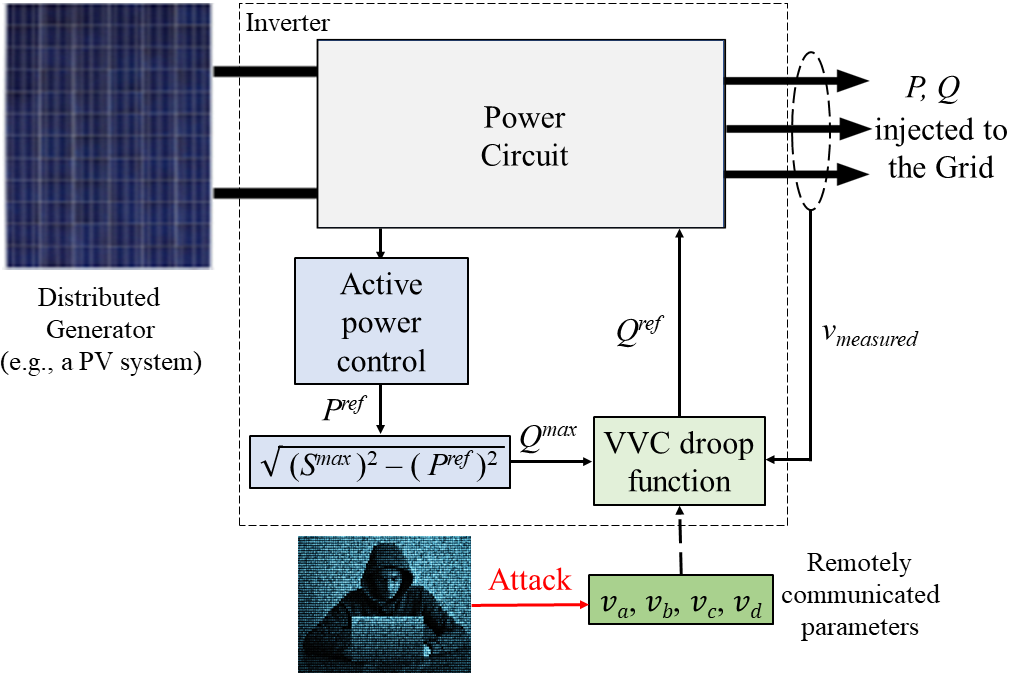}
\caption{  Block diagram of the VVC function in a smart inverter that interfaces a Distributed Generator with the grid.  Malicious entities can manipulate the remotely-communicated VVC droop parameters.  }
\label{fig:Control}
\end{figure}

In a decentralized manner, the VVC piece-wise function determines the required reactive power to be injected by the inverter ($Q_{ref}$) based on only the magnitude of the local voltage ($v$) that is often measured at the point of common coupling of the inverter \cite{VVC_paper}. 
Fig. \ref{fig:Control} illustrates the typical block diagram of inverter control.
As illustrated, for the reactive power control performed by the VVC, firstly, the maximum available reactive power ($Q_{max}$) is determined as

\begin{equation}
Q_{max} = \sqrt{ S_{max}^2 - P_{ref}^2 }
\end{equation}

\noindent where $S_{max}$ is the rated apparent power of the inverter, and $P_{ref}$ is the inverters reference active power. 
Afterward, the desired reactive power is determined using the VVC droop control function so that

\begin{equation}
Q_{ref} (t) =                       
\begin{cases}  
    Q_{max}   & \text{ } v_d \leq v   \\
    Q_{max} . \frac{(V-V_c)}{(V_d-V_c)}   & \text{ } v_c \le v < v_d   \\
    0   & \text{ } v_b \le v < v_c   \\
    -Q_{max} . \frac{(V-V_A)}{(V_B-V_A)}   & \text{ } v_a \le v < v_b   \\
    -Q_{max}   & \text{ } v < v_a   \\
\end{cases}
\end{equation}

\noindent  where $v_a$, $v_b$, $v_c$, and $v_d$ are the remotely communicated droop settings, as shown in Fig. \ref{fig:VVC}. 
These settings shape the VVC characteristics and, in return, affect how the inverter responds to changes in the measured voltage $v$.
Typically, the steeper the curve, the faster the mitigation of voltage deviation problems,  at the cost of exposing the system to possible oscillations \cite{Anna_adaptive}.

A malicious entity can remotely attack the inverter by modifying the VVC set-points, as illustrated in Fig. \ref{fig:Control}.  
This intrusion is possible if this entity (i) intercepts the communication link between the remote control center and the inverter to be attacked, (ii) \textcolor{black}{gains access to the control center, e.g., using stolen credentials or} through a supply chain attack, or (iii) exploits the vulnerabilities in the firmware of the targeted inverter. 
This attack can also be cyber-physical, providing that the malicious entities are capable of re-dispatching the set of voltage breakpoints, i.e., $v_a$, $v_b$, $v_c$ and $v_d$, that parameterize the droop curves in Figs. \ref{fig:VVC} for a subset of DGs in the distribution system to cause voltage violations \cite{Anna_DRL_DER}. 
The success of previous attacks on smart inverters, e.g., \cite{Risks}, is a clear example of the vulnerability of current inverters to remote cyberattacks. 
Once they take command of a communication channel to the targeted inverter, the adversaries can carefully manipulate the VVC curve to induce voltage oscillations in the system.

The adversaries can manipulate the four setpoints of the VVC to construct an inverted curve, i.e., a curve whose first and last segments have positive slopes \cite{edge}, or a curve with very steep non-zero first and last segments \cite{edge,Anna_adaptive}.
Inverted curves can be easily detected since the slopes of VVC inclined segments are always negative by nature, which leaves us with attacks that manipulate the slopes of the VVC curve.

To further restrict attackers, the work in \cite{edge} proposed imposing restrictions on the line connecting $v_a$ and $v_d$. However, careful attackers can bypass these restrictions by shifting the entire curve or manipulating only $v_c$ and $v_d$.
For instance, attacks shifting the VVC setpoints can be represented as

\begin{equation}
\hat{v}^{m} = \hat{v}^{pre}   + \hat{v}^{\alpha}
\end{equation}

\noindent where \textcolor{black}{ 
$\hat{v}^{m}$ is the vector of manipulated parameters, 
$\hat{v}^{pre}$ is the vector of the stable pre-attack parameters, and
$\hat{v}^{\alpha}$ is the attack vector, the amount by which the droop setpoint is manipulated.} A demonstration is provided below.

\subsection{Cyberattacks on Distributed Volt-Var Control Function}

\begin{figure*}[t!]
\centering
\includegraphics[width=0.71\linewidth,keepaspectratio]{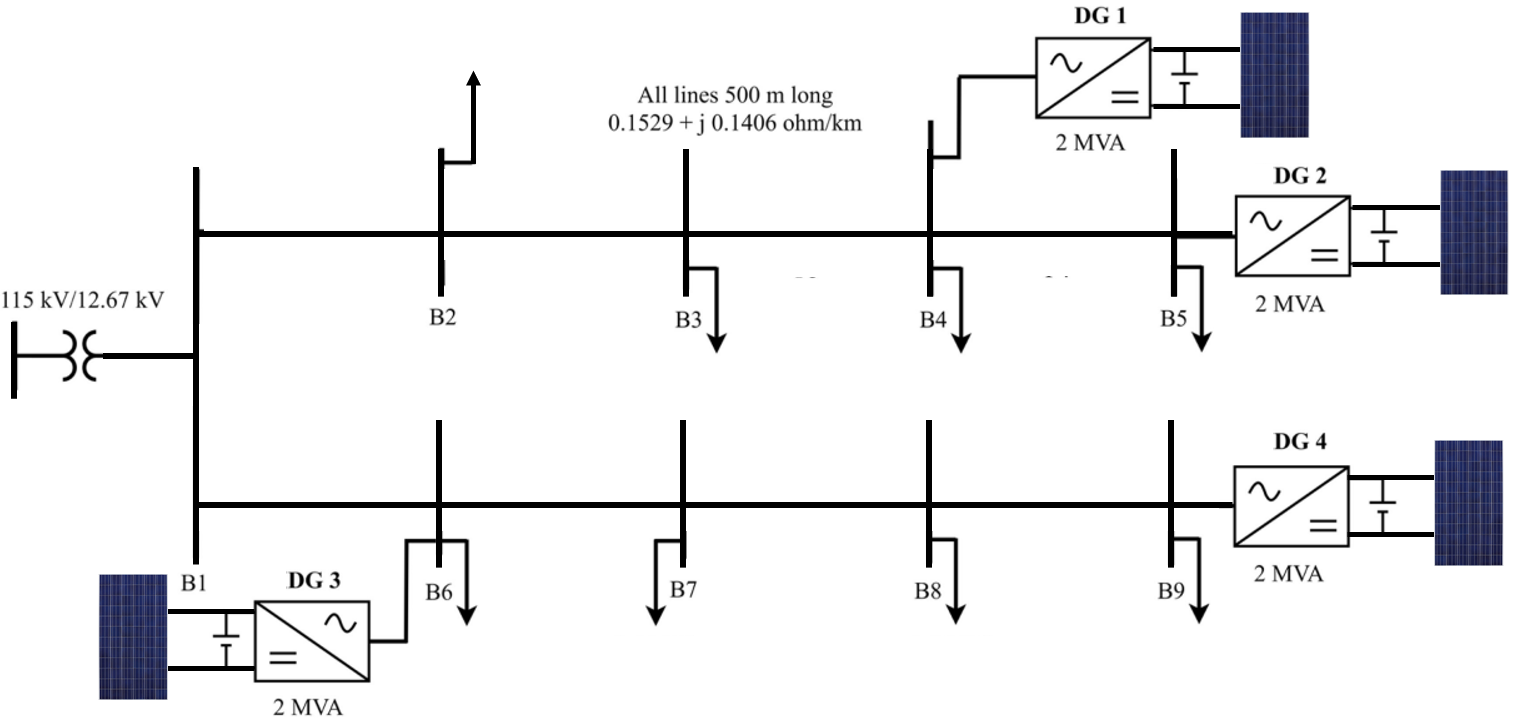}
\caption{ \textcolor{black}{Medium-voltage 9-bus part of the Canadian benchmark network equipped with four inverter-based distributed generators operating in Volt-VAr mode. } }
\label{fig:testsystem}
\end{figure*}

\begin{figure}[t!]
\centering
\includegraphics[width=0.8\linewidth,keepaspectratio]{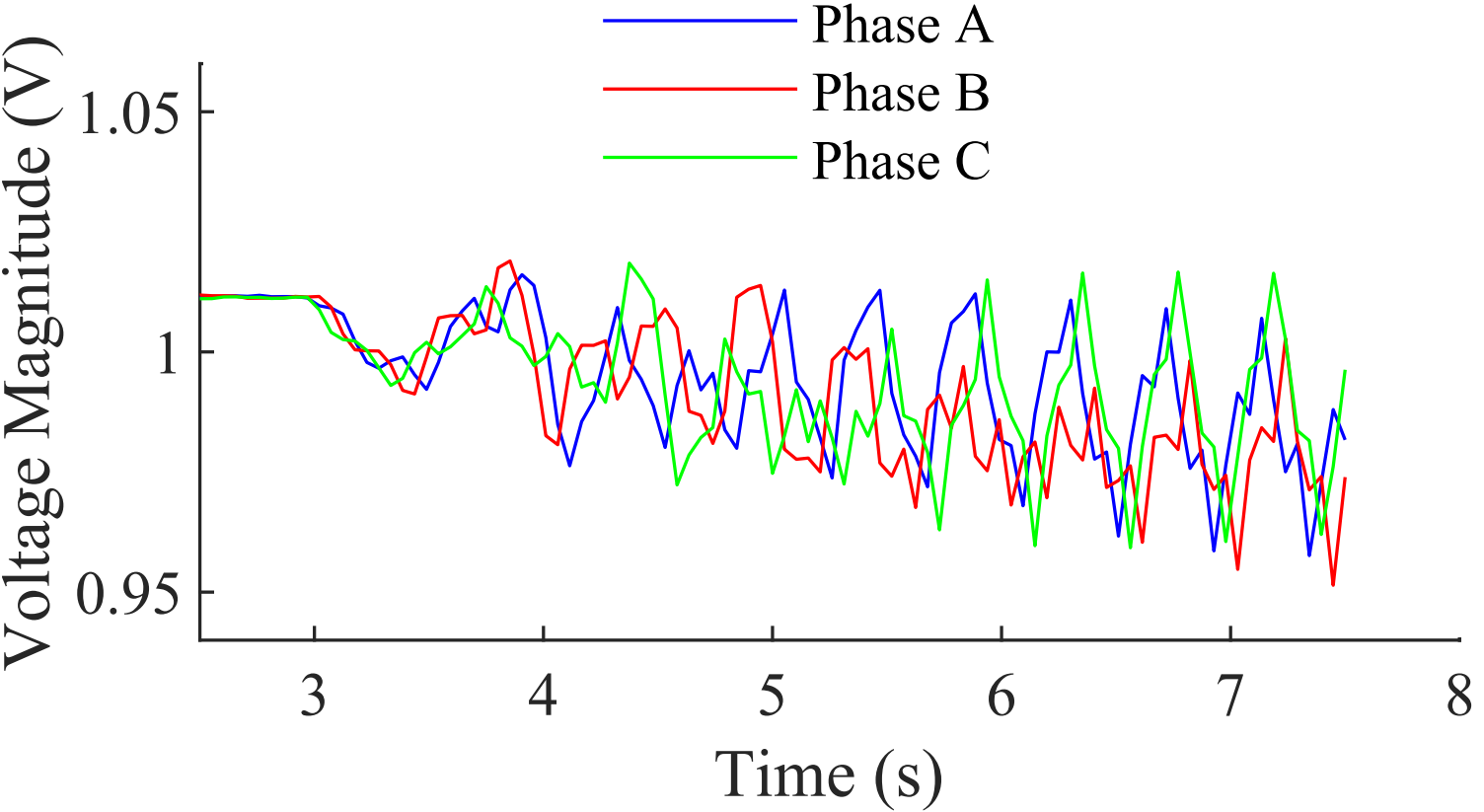}
\caption{ Oscillations in the voltage magnitude measured near DG1 following an undetected cyberattack on the VVC setpoints of the inverter of DG1.}
\label{fig:Case_attack}
\end{figure}

To demonstrate cyberattacks targeting inverters operating in VVC modes, the test system, shown in Fig. \ref{fig:testsystem}, is simulated in PSCAD/EMTDC environment.
The depicted 9-bus Canadian urban benchmark distribution network \cite{testsystem1,testsystem2} consists of two feeders, each with a rating of 8.7 MVA and an impedance of 0.1529 + j 0.1406 $\Omega$/km. The utility supplies power to these feeders through a 20-MVA 115 kV/12.47 kV transformer with a short-circuit level of 500 MVA and an X/R ratio of 6.
This system has four 2-MVA inverter-based Distributed Generators (DGs), located at buses 4, 5, 6, and 9, and interfaced with the system via 12.47 kV/0.6 kV transformers.  
DG1, which is used in this paper to demonstrate the proposed solution, is operating in VVC mode with $v_a$ =0.95, $v_b$ = 0.98, $v_c$ = 1.02 and $v_d$ =1.05, all expressed in per-unit.  Normally, $v_1$ (the local voltage of  DG1) equals  1.011 pu.
At time = 3s,  attackers, following Equation (3), commanded DG1 to update its $v_b$ and $v_c$ with the values 1.02 pu and 1.04 pu, respectively.
As a result, the operating point $v_1$, now falls between $v_a$ and $v_b$, and the inverter injects reactive power unnecessarily. A few seconds later, $v_1$ oscillations manifest, as shown in Fig. \ref{fig:Case_attack}.
It can be concluded from this section that there are many ways to manipulate the VVC droop parameters, which motivates developing an intrusion detection method to detect various malicious curve parameters, as explained in Section \ref{Section:IDM}.

\section{Developing an Intrusion Detection Method for Distributed Volt-Var Control Schemes} \label{Section:IDM}

\begin{figure}[t!]
\centering
\includegraphics[width=1.0\linewidth,keepaspectratio]{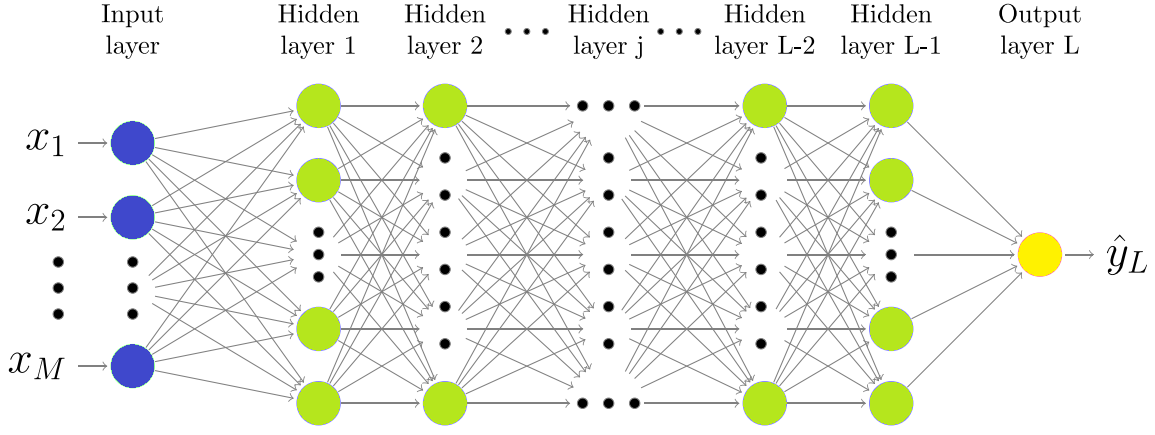}
\caption{Architecture of MLP.}
\label{fig:MLP}
\end{figure}

\textcolor{black}{The previous section demonstrated the vulnerability of VVC in smart inverters to cyberattacks. It was also shown that there are numerous ways attackers can manipulate the curve parameters. What makes this kind of cyberattack more challenging is the inherited non-linearity in the VVC control function, mainly since the inverter relies only on local measurements.
To this extent, we must develop a method to} differentiate between valid and malevolent curves once they are communicated to the inverter. 
This method must only rely on the local measurements of the inverter, i.e., the current and voltage measurements. 
\textcolor{black}{In this paper, we leverage learning-based techniques to differentiate between malevolent and legitimate settings.
The} selected model should be independent of probabilistic information, such as the specific probability density functions associated with faults or cyber-attacks. Instead, it should provide the necessary decision through only the training process.   
To achieve this goal, the proposed solution in this paper employs a Multi-Layer Perceptron (MLP) since MLPs are known for their accuracy and speed \cite{MLPref}. 
MLP architecture is illustrated in Fig. \ref{fig:MLP}. 
In this paper, the MLP is trained on features extracted from local inverter measurements to predict whether the received curve, when applied, will result in oscillations or not.

\subsection{A MultiLayer Perceptron for Detection of Malicious Curves}

To differentiate between benign  and malicious curves, the MLP maps the vector of input features $\textbf{x}$

\begin{equation}
\textbf{x} = [x_1, x_2,...,x_M]
\end{equation}

\noindent where $M$ is the number of input features, to a specific label $\hat{\textbf{y}}$, which is either 0 (for stabilizing curves) or 1 (for manipulated curves). 
To perform this classification, the MLP learns the representation of large data structures, i.e., during training, using the backpropagation concept \cite{MLPref}.
In each feedforward iteration, $x$ propagates from the input layer to the output layer passing through $L$ dense layers, as shown in Fig. \ref{fig:MLP}.

Let  $z_j$ be the input of $j^{th}$ hidden layer, where $j \in  \{1, 2, ... L\}$. Therefore, $z_j$ is the weighted sum of the outputs of the preceding layer, which can be described as

\begin{equation}
z_j =                       
\begin{cases}  
     w_1^T . \textbf{x}     + bj 
    & \text{ } j =1   \\
    w_j^T . \hat{\textbf{y}}_{j-1}     + b_j 
    & \text{ } j \in  \{2,...,L\}
\end{cases}
\end{equation}

\noindent where $w_j$ and $b_j$ are the weights vector and bias vector of $z_j$, respectively. Accordingly, $ \hat{\textbf{y}}_{j-1}  $ is the output of the $(j-1) ^{th}$ layer. 
In detail, the output of layers 1 to $L-1$ is obtained as

\begin{equation}
\hat{\textbf{y}}_j =  \text{max}(0,z_j )
\end{equation}

Finally, the output of the final layer ($L^{th}$ layer)  is

\begin{equation}
\hat{\textbf{y}}_L =  \frac{1}{1+ e^{-z_L}}  
\end{equation}

\noindent which is obtained using the sigmoid function and yields a binary output, i.e.,  1 for a cyberattack or 0 for a benign curve.

\subsection{Automatic Hyperparameter Tuning of the MLP}

The goal of the training phase is to obtain the MLP model with the best hyperparameter values that result in the highest accuracy. Hyperparameters utilized include the number of layers, the size of each layer (number of nodes), the weights, the regularization strength ($\lambda$), and the selection of the activation function for each node.  
This paper leverages the usage of automatic MLP tuning to obtain the desired MLP model in a systematic way. Here, the Random Search algorithm is used to update the hyperparameter values \cite{randomsearch}. 
Here, the Random Search algorithm aims to maximize the MLP's accuracy
using the hyperparameters as variables.

\section{Simulation Results} \label{Section:Case} 

This section depicts how the MLP, utilized by the proposed intrusion detection system, is trained and tested to detect cyberattacks that would otherwise disrupt the stability of the system as demonstrated in Section \ref{Section:VVC}. A.

\subsection{Data Generation and Features for Training the MLP}

To ensure that the proposed solution is truly capable of detecting malicious curves, a wide range of scenarios are simulated. Using DG1 in the test system shown in Fig. \ref{fig:testsystem}, \textcolor{black}{2000 malicious scenarios are simulated, during which destabilizing curves are communicated to the inverter of DG1. To generate the scenarios, the voltage setpoints $\hat{v}$ are varied following Equation (3), under different system loading conditions. Similarly,  2000 legitimate scenarios are simulated where stabilizing curves are applied.}
For practicality, all the simulated curves have non-zero segments.
In each scenario, the inverter's three-phase current and voltage measurements are recorded from the instant the new VVC is received until 10 seconds later. 
From these measurements, a set of features is extracted to train the MLP from each sample, which includes: (i) new VVC curve parameters,
(ii) magnitudes of the three-phase voltage and current phasors, (iii) the direct and quadrature currents and voltages, i.e., 3-phase voltages and currents after being transformed to the direct-quadrature (\textit{d-q}) domain, which are already calculated by smart inverters \cite{VVC_paper}, 
and (iv) $\zeta$, which is an intuitive \textcolor{black}{reflection} of the oscillations in each current/voltage magnitude waveform, \textcolor{black}{defined} as

\begin{equation}
 \zeta  = c \textcolor{white}{.} |v_{pcc} - vn| ^p
\end{equation}

\noindent where $| . |$ denotes the absolute value of the enclosed argument,  
$v_{pcc}$ is the root mean squared voltage measured at the point of common coupling of the inverter,
$v_n$ is the nominal value of $v_{pcc}$ per-unit, and $c$ and $p$ are positive integer constants.
When the VVC curve is legitimate, it results in a steady-state condition, where the difference between $v_{pcc}$ and $v_n$ is zero.
However, following the usage of a malicious curve, oscillations in $v_{pcc}$ manifest and $v_{pcc}$ can either increase or fall below $v_n$. 
$p$ and $c$ help magnify the amplitude of these oscillations, thus easing the differentiation between minor fluctuations that naturally exist in the power system and oscillations resulting from malicious VVC curve parameters. This paper implements $p$ and $c$ as 2 and 100, respectively. 
%
%
%
Table \ref{table:features} summarizes the utilized features. 
In each scenario, the above features are inputted twice to the MLP, i.e., the features' values before and after the new curve is applied.  
Afterward, the generated dataset is labeled as either `malicious' or `legitimate,' where malicious curves are those that would eventually result in oscillations appearing in the voltage of DG1, and legitimate curves are the rest of the curves.

\begin{table}[t]
\centering
\begingroup
\caption{  Summary of Utilized Features}
\textcolor{black}{
\begin{tabular}{c | c }\hline
\centering
\makebox{Features }
&
\makebox{Remarks}
\\   \hline 
\rule{0pt}{2ex}    
$v_a, v_b, v_c, v_d $  & Parameters of the new VVC curve
\\
\rule{0pt}{3ex}    
$|I_1|, |I_2|, |I_3|, $  & Magnitudes of local current measurements
\\
\rule{0pt}{3ex}    
$|V_1|, |V_2|, |V_3|, $  & Magnitudes of local voltage measurements
\\
\rule{0pt}{3ex}    
$I_d, I_q, V_d, V_q,$  &  Currents and voltages in {dq} domain   
\\
\rule{0pt}{3ex}    
$\zeta_{V_1}$, $\zeta_{V_2}$, $\zeta_{V_3}$  &  Indication of oscillations in the voltage magnitudes 
\\
 \hline 
\end{tabular}
}
\label{table:features}
\endgroup
\end{table}

\subsection{Training of the MLP}

The dataset generated above is shuffled and then randomly split into 80\% for training and 20\% for testing. 
The training portion of the dataset is used to train the MLP, which is performed in MATLAB environment. 
When the training commences, hyperparameters are randomly initialized.
To obtain a more reliable estimate of the MLP model's performance,
\textcolor{black}{training is performed in conjunction with k-fold cross validation \cite{kfcv}.
That is,} in each epoch of the training, the training dataset is divided into five equal folds, and then five separate models are trained (each model is trained on a different fold and validated against the remaining four folds).   
In this paper, the optimum MLP model comes with three layers that consist of 7 nodes, 159 nodes, and 1 node, respectively. The layers are ReLU-activated and $\lambda$ =  $5.2128 \times 10^{-7}$.

\subsection{Evaluation Metrics}

A successfully detected malicious curve is labeled as a true positive (\textit{T}P) case. Correctly predicted legitimate curves are considered true negative (\textit{TN}) cases. Accordingly, false positive (\textit{FP}) and false negative (\textit{FN}) cases are mispredicted normal curves and malicious curves, respectively. To corroborate the performance of the proposed intrusion detection system, the following metrics are used:

\begin{equation}
\text{\textit{Accuracy}} = \frac{{\text{\textit{TP + TN}}}}{{\text{\textit{TP + TN + FP + FN}}  }}
\end{equation}
\begin{equation}
\text{\textit{Precision}} = \frac{{\text{\textit{TP}}}}{{\text{\textit{TP}} + \text{\textit{FP}}}}
\end{equation}
\begin{equation}
\text{\textit{Recall}} = \frac{{\text{\textit{TP}}}}{{\text{\textit{TP}} + \text{\textit{FN}}}}
\end{equation}
\begin{equation}
\text{\textit{F1-score}} = 2 \times \frac{{\text{\textit{Precision}} \times \text{\textit{Recall}}}}{{\text{\textit{Precision}} + \text{\textit{Recall}}}}
\end{equation}

\subsection{Evaluation Results}

The trained/optimum MLP model \textcolor{black}{is then tested on the previously unseen} 20\% of the dataset held out for testing. 
Table \ref{table:results} depicts the performance evaluation results of the proposed intrusion detection method. 
Overall, 99.7\% of malicious VVC curves are detected,
which reflects the ability of the proposed solution to protect the inverter-system interaction from undesirable oscillations. 
Nonetheless, the rest of the inverters can still mitigate these undetected cases, e.g., using the approach described in \cite{Anna_adaptive}. 
On the other hand,  99.8\% of legitimate VVC curves are correctly classified and therefore permitted, which shows that the proposed method minimally restricts the controllability of the VVC process and system operator. 

\begin{table}[t]
\centering
\begingroup
\caption{  Performance of the Proposed intrusion detection method}
\begin{tabular}{c | c }\hline
\centering
\makebox{Metric }
&
\makebox{Reults }
\\ 
\hline
\rule{0pt}{2ex}    
\textit{Accuracy} & 99.75\% \\
\rule{0pt}{2ex}    
\textit{Precision} & 99.8 \%   \\
\rule{0pt}{2ex}    
\textit{Recall} & 99.7 \%   \\
\rule{0pt}{2ex}    
\textit{F1-score} & 99.74 \%  \\ 
\hline
\end{tabular}
\label{table:results}
\endgroup
\end{table}

\begin{table}[t]
\centering
\begingroup
\caption{ Confusion Matrix}
\hspace{50pt} \textit{Predicted Scenario } 
\vspace{3pt}
\\
\begin{tabular}{c | c| c}\hline
\centering
\makebox{ \textit{ True Scenario} }
&\makebox{ Malicious  }
&\makebox{ benign  }
\\\hline
\rule{0pt}{2ex} Malicious  & 99.7 \%  & 0.2\%
\\ 
\rule{0pt}{2ex} benign   & 0.3\% &  99.8 \%  
\\
\hline
\end{tabular}
\label{table:confusionmatrix}
\endgroup
\end{table}

 \section{Securing Volt-Var Control Against Setpoint Manipulation Cyberattacks} \label{Section:flowchart}

After integrating the proposed solution within smart inverters,
once a new curve is received, it will be evaluated by the proposed solution, which will determine if it is a stabilizing curve, and thus should be permitted or a destabilizing curve, and thus the VVC continues adopting the old set of setpoints.
Implementing the proposed scheme helps prevent many oscillations before they happen on the system. 
\textcolor{black}{When a malicious curve is detected,} there are a few options regarding how the attacked inverter should respond \cite{edge}. For instance, the power system operator can pre-engineer all inverters in the system to turn off the VVC mode and switch to the unity-power-factor mode once attacked. This approach will prevent attackers from threatening the voltage stability of the power system. However, increased active power injection by the DGs can result in increased system losses.
Alternatively, the utility may require attacked inverters to self-isolate. Nonetheless, this option opens doors to denial-of-service attacks.
Furthermore, attacked inverters can revert to the last communicated stable curve or switch to a flattened curve pre-designed as a backup curve \cite{Anna_adaptive}. 
Generally, more research is required in this direction.
\textcolor{black}{It is also interesting to investigate the possibility of utilizing one-class classifiers, as opposed to the considered binary MLP. In addition, investigating the interpretability of the utilized classifier is another interesting research direction.}

 \section{Conclusion} \label{Section:conclusion} 

Distributed VVC is a widely adopted control mode for inverters equipping DGs. 
Smart inverters are required to support remote updating of the VVC parameters, which poses a vulnerability that malicious entities can exploit and maliciously manipulate the VVC curves of smart inverters and, in return, threaten the distribution system's voltage stability. 
This paper presented an intrusion detection method that can prevent malicious cyberattacks targeting the setpoints of the VVC mode.
When the inverter receives a new set of VVC curve setpoints, the proposed method \textcolor{black}{uses an MLP to confirm that these curves are not malicious or will result in oscillations in the system voltage.}
Only legitimate stabilizing curves will be allowed. Otherwise, the inverter keeps using the last known stabilizing curves.
The proposed method is entirely decentralized, utilizes only the local measurements available for inverters, and requires no additional communication. 
The performance of the proposed solution was verified using the 9-bus Canadian Urban benchmark system equipped with inverter-based DGs and simulated in PSCAD/EMTDC. 
The training of the MLP was performed in a MATLAB environment.  
Our results show that the proposed intrusion detection method can accurately differentiate between malicious and legitimate stabilizing  VVC curves. Therefore, the proposed method is recommended for integration with smart inverters.

\section*{Acknowledgment}
This work is supported by CIRA-013-2020, Khalifa University, UAE, and VRI20-07, ASPIRE Virtual Research Institute Program, Advanced Technology Research Council, UAE.

\ifCLASSOPTIONcaptionsoff
  \newpage
\fi

\end{document}